\documentclass[showpacs,amsmath,amssymb]{revtex4}
\input{epsf}

\begin{document}

\title{\bf Tachyons, Scalar Fields and Cosmology}

\author{Vittorio Gorini}
\affiliation{Dipartimento di Scienze Fisiche e Matematiche,
Universit\`a dell'Insubria, Como, Italy \\
and INFN, sez. di Milano}

\author{Alexander Kamenshchik}
\affiliation{Dipartimento di Scienze Fisiche e Matematiche,
Universit\`a dell'Insubria, Como, Italy \\
and L.D. Landau Institute for Theoretical Physics of the Russian
Academy of Sciences, Moscow, Russia}

\author{Ugo Moschella}
\affiliation{Dipartimento di  Scienze Fisiche e Matematiche,
Universit\`a dell'Insubria, 22100 Como, Italy \\
and INFN, sez. di Milano}

\author{Vincent Pasquier}
 \affiliation{Service de Physique Th\'eorique, C.E. Saclay, 91191
Gif-sur-Yvette, France}
 \begin{abstract} We study the role that tachyon fields
may play in cosmology as compared to the well-established use of
minimally coupled scalar fields. We first elaborate on a kind of
correspondence existing between tachyons and minimally coupled
scalar fields; corresponding theories give rise to the same
cosmological evolution for a particular choice of the initial
conditions but not for any other. This leads us to study a
specific one-parameter family of tachyonic models based on a
perfect fluid mixed with a positive cosmological constant. For
positive values of the parameter one needs to modify Sen's action
and use the $\sigma$-process of resolution of singularities. The
physics described by this model is dramatically different and much
richer than that of the corresponding scalar field. For particular
choices of the initial conditions the universe, that does mimick
for a long time a de Sitter-like expansion, ends up in a finite
time in a special type of singularity that we call a {\em big
brake}. This singularity is characterized by an infinite
deceleration.
\end{abstract}
 \maketitle
\section{Introduction}
The recently discovered cosmic acceleration
\cite{accel1,accel2,accel3} has put forward the problem of
unravelling the nature of the so called dark energy responsible
for such phenomenon (for a review see \cite{Star-Varun,Pad-rep}).
The crucial feature of the dark energy which ensures an
accelerated expansion of the universe is that it breaks the strong
energy condition.

The tachyon field arising in the context of string theory
\cite{Sen,Sen1,Sen2,Garousi} provides one example of matter which
does the job. The tachyon has been intensively studied during the
last few years also in application to cosmology
\cite{Gibbons1}--\cite{bilbao1}; in this case one usually takes
Sen's effective action \cite{Sen} for granted and studies its
cosmological consequences without worrying about the
string-theoretical origin of the action itself. We take this
attitude in the present paper.

However, it would be reasonable, before considering concrete
tachyon cosmological models, to answer a simple question: is the
tachyon field of real interest for cosmology? Indeed, it is
well-known that for isotropic cosmological models, for a given
dependence of the cosmological radius on time it is always
possible to construct a potential for a minimally coupled scalar
field (in brief: scalar field) model, which would reproduce this
cosmological evolution (see e.g. \cite{Star-scal}), provided
rather general reasonable conditions are satisfied.  Since a
similar statement holds also for cosmological models with
tachyons, one can find some kind of correspondence between
minimally coupled scalar field models and tachyon ones describing
the same cosmological evolution \cite{Padman}. Therefore, a
natural question arises: does it make sense at all to study
tachyon cosmological models in place of the traditional scalar
field models?

In our opinion, the point is that the correspondence between
tachyon and scalar field cosmological models is a rather limited
one and amounts to the existence of "corresponding" solutions of
the models, obtained by imposing certain special initial
conditions. If one moves away from these conditions the dynamics
of the tachyon model can become more complicated and very
different from that of its scalar field cousin.

In this paper we consider some examples of scalar and tachyon
field isotropic cosmological models having coincident exact
solutions. All the exact solutions considered here actually arise
as solutions of some isotropic perfect fluid cosmological models.

Some of the tachyon and scalar field potentials that we consider
here are well known in the literature and are widely used as
quintessential models. We will introduce also a new (at least to
our knowledge) tachyon model that is based on the cosmic evolution
driven by the mixture of a cosmological constant and of a fluid
with equation of state $p = k \varepsilon$, with $ -1 < k \leq 1$
($\varepsilon$ and $p$ denote as usual the energy density and the
pressure respectively). The corresponding tachyon potential $V(T)$
is represented by a rather cumbersome trigonometric function of
the tachyon field. When the parameter $k$ is positive the
potential becomes  imaginary passing through zero for finite
values of the tachyon field $T$. Thus, it might seem that these
features would kill the model; on the contrary, using the methods
of the qualitative theory of ordinary differential equations, it
is possible to extract from the model a  dynamics which is
perfectly meaningful also for positive $k$. We find that this
dynamics is quite rich. In particular, when the value of the
parameter $k$ is positive, the model possesses two types of
trajectories: trajectories describing an eternally expanding
universe and those hitting a cosmological singularity of a special
type that we have chosen to call a {\em big brake}. Thus, starting
with a simple perfect fluid model and trying to reproduce its
cosmological evolution in scalar field and tachyon models, we
arrive to a relatively simple scalar field potential with a
correspondingly simple dynamics and to a complicated tachyon
potential. The latter provides us with a model having a very
interesting dynamics giving rise to very different cosmological
evolutions and opening in turn opportunities for some non-trivial
speculations about the future of the universe.

The structure of the paper is as follows. In the second section we
study the problem of the correspondence between tachyon and scalar
field cosmological models and give some examples in the third
section. In the fourth section we suggest a way to go beyond the
limitations of tachyonic models and we make use of this idea in
the fifth and sixth sections where we study the dynamics and the
cosmology of a particular (toy) tachyon model introduced in Sec.
3. In the last section we conclude our paper by some speculations
on the future evolution of the universe arising from the analysis
of our model.

\section{The correspondence}\label{uu}

We consider a flat Friedmann cosmological model $ds^2 = dt^2 -
a^2(t)dl^2$ of a universe filled with some perfect fluid and
suppose that the cosmological evolution $a=a(t)$ is given; the
Friedmann equation
\begin{equation}
\frac{\dot{a}^2}{a^2} = \varepsilon  \label{Friedmann1}
\end{equation}
provides the dependence $\varepsilon = \varepsilon(t)$ of the
energy density of the fluid on the cosmic time (we have put $8\pi
G/3 = 1$ for convenience). Then, the equation for energy
conservation
\begin{equation}
\dot{\varepsilon} = - 3 \frac{\dot{a}}{a} (\varepsilon + p)
\label{en-cons}
\end{equation}
fixes the pressure $p=p(t)$; therefore an equation of state $p =
p(\varepsilon)$ can in principle be written describing the unique
fluid model compatible with the given cosmic evolution (provided
that $\dot{\varepsilon} \not= 0$).

Now, let us suppose  that the matter content of the universe is
modeled by a homogeneous tachyon field $T(t)$ described by Sen's
Lagrangian density \cite{Sen,Sen1}:
\begin{equation}
L = -V(T)\sqrt{1 - {\dot T}^2}. \label{tachyon}
\end{equation}
The energy momentum tensor is diagonal and  the corresponding
field-dependent  energy density and pressure are given by
\begin{equation}
\varepsilon = \frac{V(T)}{\sqrt{1-\dot{T}^2}}, \label{energy}
\end{equation}
\begin{equation}
p = -V(T)\sqrt{1-\dot{T}^2}, \label{pressure}
\end{equation}
while the field equation for the tachyon is written as follows
\begin{equation}
\frac{\ddot{T}}{1-\dot{T}^2} + \frac{3\dot{a}{\dot{T}}}{a} +
\frac{V_{,T}}{V} = 0. \label{tac-eq}
\end{equation}
One can try and find a potential $V(T)$ so that, for certain
suitably chosen initial conditions on the tachyon field, the scale
factor of the universe is precisely the given $a(t)$. A similar
construction can be attempted for a minimally coupled scalar
field, described by the Lagrangian density
\begin{equation}
L = \frac12\dot{\varphi}^2 - U(\varphi). \label{scalar}
\end{equation}
A given cosmological evolution $a = a(t)$ can therefore be used to
establish a sort of correspondence between the two potentials
$V(T)$ and $U(\varphi)$ (see also \cite{Padman}).

Let us see in some detail how this works and what the sought
correspondence really means. It is often more practical to use the
scale factor $a$ to parameterize the cosmic time; this can be done
provided $\dot a \not = 0$.
From Eqs. (\ref{energy}) and
(\ref{pressure}) it follows that
\begin{equation}
{\dot T}^2 =\frac{p+\varepsilon}{\varepsilon}.
\end{equation}
By using Eqs. (\ref{Friedmann1}) and (\ref{en-cons}) the latter
equation can be rewritten as follows
\begin{equation}
T' = \frac{1}{a}\sqrt{\frac{(\varepsilon + p)}{\varepsilon^2}} =
\frac{1}{a}\sqrt{\frac{-\varepsilon'a}{3\varepsilon^2}}
\label{Tdef}
\end{equation}
where ``prime'' denotes the derivative with respect to  $a$. Once
$a(t)$ and therefore $\varepsilon(a)$ is specified (see Eq.
(\ref{Friedmann1})), Eq. (\ref{Tdef}) can be integrated  to give
\begin{equation}
T = \Phi(a) = \int^a\frac{dx}{x}\sqrt{\frac{-\varepsilon'(x)\, x
}{3\varepsilon^2(x)}} \label{Phidef}
\end{equation}
(an arbitrary additive integration constant is hidden in the
unspecified integration limit). By inverting Eq.(\ref{Phidef}) $a
= \Phi^{-1}(T)$ and making use of the relation
\begin{equation}
V = \sqrt{-\varepsilon p} =
\sqrt{\frac{(\varepsilon^2a^6)'}{6a^5}}, \label{Tpot}
\end{equation}  the shape of the required tachyon potential
$ V = V(a) = V(\Phi^{-1}(T))$ can be found.

For the minimally coupled scalar field we have similar formulas.
The analogs of Eqs. (\ref{energy}) and (\ref{pressure}) for the
scalar field are
\begin{equation}
\varepsilon = \frac12\dot{\varphi}^2 + U(\varphi),\label{energys}
\end{equation}
\begin{equation}
p = \frac12\dot{\varphi}^2 - U(\varphi). \label{pressures}
\end{equation}
From these we get
\begin{equation}
\varphi' = \frac{1}{a}\sqrt{\frac{(\varepsilon + p)}{\varepsilon}}
= \frac{1}{a}\sqrt{\frac{-\varepsilon'a}{3\varepsilon}}
\label{sdef}
\end{equation}
and
\begin{equation}
U = \frac{1}{2}(\varepsilon - p) = \frac{(\varepsilon
a^6)'}{6a^5}. \label{Udef}
\end{equation}
Integration of Eq. (\ref{sdef}) gives
\begin{equation}
\varphi  = F(a) = \int^a\frac{dx}{x}\sqrt{\frac{-\varepsilon'(x)\,
x}{3\varepsilon(x)}} \label{Fdef}
\end{equation}
and as before $U = U(a)=U(F^{-1}(\varphi))$. The formulas above
establish a kind of correspondence between the potentials $U$ and
$V$ in the sense that the cosmologies resulting from such
potentials admit the given cosmological evolution $a = a(t)$ for
suitably chosen initial conditions on the fields. However, it has
to be stressed that for arbitrary initial conditions on the fields
the cosmological evolutions (within the corresponding models) may
be drastically different. Furthermore, changing the initial
conditions, say, for a minimally coupled scalar field theory, one
gets different cosmological evolutions which, in turn, can be
reproduced in entirely different tachyon theories and viceversa.
Thus, any scalar field potential has a whole (one-parameter)
family of corresponding tachyon potentials and the same is true
the other way around.

 We can learn something more
by expressing the fields $T$ and $\varphi$ and the potentials
$V(T)$ and $U(\varphi)$ in terms of the Hubble variable $h =
\dot{a}/a$. Since
\begin{equation}
\dot{h} = -\frac32(\varepsilon + p) \label{hdot}
\end{equation}
one has easily that
\begin{equation}
\dot{T}^2 = \frac{\varepsilon + p}{\varepsilon} =
-\frac{2\dot{h}}{3h^2}, \;\;\;\;\;\; V(T) =
\sqrt{h^2\left(\frac23\dot{h} + h^2\right)}, \label{Vpot}
\end{equation}
and that
\begin{equation}
\dot{\varphi}^2 = \varepsilon + p = -\frac23\dot{h},\;\;\;\;\;\;
U(\varphi) = h^2 + \frac13 \dot{h}. \label{Upot}
\end{equation}
Eqs. (\ref{Vpot}) and (\ref{Upot}) call for the condition $
\varepsilon + p \geq 0,$ or equivalently $\dot{h} \leq 0$. For the
tachyon model Eq. (\ref{Vpot}) requires the additional condition
\begin{equation} \dot{h} \geq -\frac32 h^2
\label{realizable2}
\end{equation}
which is equivalent to ask for a negative pressure $p \leq 0$. It
seems therefore that tachyonic models are more restrictive than
minimally coupled scalar fields in their possibility to describe
cosmological evolutions. We will see in an example that this
seemingly negative conclusion can be overcome and pleasant
surprises may arise.

\section{Examples}
We now illustrate the above considerations by means of some
explicitly solvable examples.  In these examples we assume, as
usual, that the initial moment of time, $t = 0$, corresponds to
the cosmological singularity. All these examples are based on
simple models of perfect fluids as specified by their equations of
state.

\begin{enumerate}
\item The starting point of the first example is a model of
universe filled with a perfect fluid with equation of state $
p = k\varepsilon$, with $ -1 < k \leq 1$; 
the Hubble variable of this model is given by
\begin{equation}
h(t) = \frac{2}{3(1+k)t}.
\label{Hubblemod1}
\end{equation}
A minimally coupled scalar field theory that produces the same
evolution $h(t)$ for suitably chosen initial conditions can be
based on one of the following scalar potentials, which are
obtained by the procedure outlined in the previous section:
\begin{equation}
U_\pm(\varphi) = \frac 29\,  \frac{1-k}{(1+k)^2}\ e^{\pm
3\sqrt{1+k}(\varphi-\varphi_0)};\label{scalpot1}
\end{equation}
the limiting case $k=1$ gives a minimally coupled massless field.
For $k>1$ the potential becomes negative reflecting the fact that
the velocity of sound is bigger than the speed of light.

Both choices of sign in the exponent are acceptable and the
normalization of the potentials can be chosen arbitrarily (the
constant $\varphi_0$); then, the exact solutions of the field
equations providing the prescribed cosmological evolution
(\ref{Hubblemod1}) are, respectively, the following:
\begin{equation}
\varphi_\pm(t) = \mp \frac{2}{3\sqrt{1+k}}\ln t +
\varphi_0.\label{incond1}
\end{equation}
The theories are obviously connected by the symmetry operation $
\varphi - \varphi_0 \rightarrow -(\varphi - \varphi_0)$, which
exchanges the potentials and the corresponding solutions.

To construct a tachyonic field theory we have to restrict our
attention to the case $p<0$, i.e. $k<0$. Following the same
procedure we get a tachyon field model, based on the potential
\begin{equation}
V(T) = \frac{4}{9} \ \frac{\sqrt{-k}}{(1+k)}\ \frac{1}{(T-T_0)^2}.
\label{V1}
\end{equation}
There are again two exact solutions of the field equations
(\ref{Hubblemod1}), which now do co-exist within the same model:
\begin{equation} T_\pm(t) = \pm \sqrt{1+k}\ t +
T_0. \label{incond4}
\end{equation}

This tachyonic model has been studied in \cite{Feinstein,Padman},
where also the correspondence with a minimally coupled scalar
field theory with exponential potential was noticed. There exists
a vast literature on the latter (see e.g.
\cite{matarrese}-\cite{Napoli}).

\item Now we add a positive cosmological constant to the previous
model, i.e. we consider the mixture
\begin{equation}
 p_1 = k\varepsilon_1,\ -1 < k \leq 1,\,\,p_\Lambda = -\varepsilon_\Lambda =
 -\Lambda, \ \ \ \Lambda>0.
\label{model2}
\end{equation}
This is the same as one fluid with state equation
\begin{equation}
p = k\varepsilon - (1+k)\Lambda. \label{model21}
\end{equation}
The evolution of the Hubble variable is now given by
\begin{equation}
h(t) = \sqrt{\Lambda}\coth\frac{3\sqrt{\Lambda}(1+k)t}{2}.
\label{Hubblemod2}
\end{equation}
The corresponding minimally coupled scalar field theory is based
on the following potential:
\begin{equation}
U(\varphi) = \Lambda\left(1 + \frac{1-k}{2}\sinh^2
\frac{3\,\sqrt{1+k}\,(\varphi- \varphi_0)}{2}\right). \label{U2}
\end{equation}
There are two exact solutions of the field equations reproducing
the given cosmic evolution (\ref{Hubblemod2}):
\begin{equation}
\varphi_\pm(t) = \pm \frac{2}{3\sqrt{1+k}}\ \ln \ \tanh
\frac{3\sqrt{\Lambda}(1+k)t }{4} + \varphi_0. \label{incond6}
\end{equation}
To find the tachyonic model we observe first of all that to have
the restriction $p<0$ satisfied it is sufficient to require that
$-1<k\leq 0$; this condition is also necessary if one demands that
$p<0$ during all the stages of the cosmological evolution and for
any choice of the initial conditions in the tachyonic model. With
this condition we obtain the following more complicated tachyon
potential:
\begin{equation}
V(T) =
\frac{\Lambda}{\sin^2\left[\frac{3}{2}{\sqrt{\Lambda\,(1+k)}\
(T-T_0)}\right]} \sqrt{1 - (1+k)
\cos^2\left[\frac{3}{2}{\sqrt{\Lambda\,(1+k)}\,(T-T_0)}\right]}.
\label{V2}
\end{equation}
Still, the corresponding exact solutions can be found and are
given by
\begin{equation}
T(t) = \pm\frac{2}{3\sqrt{\Lambda(1+k)}}\arctan
\sinh\frac{3\sqrt{\Lambda}(1+k)t}{2} + T_0. \label{incond8}
\end{equation}
For $k>0$ there are values of the tachyon field for which the
potential (\ref{V2}) becomes imaginary passing through zero.

When $\Lambda$ tends to zero one expects to recover the model
studied in the  first example. This is indeed the case for the
tachyonic field.

As for the scalar potential (\ref{U2}) the situation is a bit
trickier: the correct limit can be obtained only by letting the
constant $\varphi_0$ vary with $\Lambda$ as follows:
\begin{equation}
\varphi_{02} = \varphi_{01} \pm \frac{1}{3\sqrt{1+k}} \ln
\frac{\Lambda(1- k)}{2}, \label{expdep}
\end{equation}
where the subscripts 1 and 2 refer to the corresponding examples.
This is because the theory has a symmetry which is lost in the
limit.

Notice that the scalar field potential (\ref{U2}) is well-defined
at $-1 < k \leq 1$ and the model possesses the exact solution
(\ref{incond6}) which corresponds to the cosmological evolution
(\ref{Hubblemod2}). The dynamics of the universe within the
tachyonic models (\ref{V1}) and (\ref{V2}) will be studied in
detail in the next sections for general initial conditions.  We
shall show that both models can be extended to the interval
$0<k\leq 1$, actually making them richer than the corresponding
scalar minimally coupled theories. \\
\item  In the third example we consider a perfect fluid whose
equation of state is as follows:
\begin{equation}
\varepsilon + p =\gamma\varepsilon^{\lambda},\, \ \ \ 0<
\gamma<1,\, \ \ \ \lambda>1. \label{model3}
\end{equation}
In this case the Hubble variable is
\begin{equation}
h(t) = \left(\frac32\, \gamma\ (2\lambda-1)\,
t\right)^{\frac{1}{1-2\lambda}}. \label{Hubblemod3}
\end{equation}The scalar potential has the form
\begin{equation}
U(\varphi) = \left(\frac{3\sqrt{\gamma}(\lambda-1)(\varphi-
\varphi_0)}{2}\right)^{-\frac{2}{\lambda-1}}-
\frac{\gamma}{2}\left(\frac{3\sqrt{\gamma}(\lambda-1)(\varphi-
\varphi_0)}{2}\right)^{-\frac{2\lambda}{\lambda-1}}. \label{U3}
\end{equation}
The exact solutions are
\begin{equation}
\varphi(t) = \pm \sqrt{\gamma}\frac{1-2\lambda}{1-\lambda}
\left(\frac32\gamma(2\lambda-1)t\right)
^{\frac{1-\lambda}{1-2\lambda}} + \varphi_0. \label{incond10}
\end{equation}
The tachyon potential is
\begin{equation}
V(T) = \sqrt{\left(\frac{3\sqrt{\gamma}\lambda(T-T_0)}{2}\right)^{-
\frac{4}{\lambda}}-\gamma\left(\frac{3\sqrt{\gamma}\lambda(T-T_0)}{2}\right)^{-
\frac{2(1+\lambda)}{\lambda}}},
\label{V3}
\end{equation}
with  the following exact solutions:
\begin{equation}
T(t) = \pm \sqrt{\gamma}\frac{2\lambda-
1}{\lambda}\left(\frac32\gamma(2\lambda-1)t\right)
^{\frac{-\lambda}{1-2\lambda}} + T_0. \label{incond12}
\end{equation}

The scalar field model with the potential (\ref{U3}) was studied
in \cite{Barrow2} while the corresponding tachyon model with the potential
(\ref{V3}) was discussed in \cite{Feinstein,Padman}. \\
\item  In our last example we consider the Chaplygin gas,
described by the following equation of state:
\begin{equation}
p = -\frac{A}{\varepsilon},\ A>0.
\label{model4}
\end{equation}
In this example the evolution $h(t)$ is given only implicitly
\cite{we} by the formula:
\begin{equation}
t = \frac{1}{6 A^{1/4}}\left(\ln\frac{h+A^{1/4}}{h-A^{1/4}} -
2\arctan \frac{h}{A^{1/4}} + \pi\right), \label{implicit}
\end{equation}
However, as shown in \cite{we}, the scalar potential can be
reconstructed using the known explicit dependence of $\varepsilon$
on $a$ and one gets
\begin{equation}
U(\varphi) = \frac{1}{2}\sqrt{A}\left(\cosh 3(\varphi-\varphi_0) +
\frac{1}{\cosh 3(\varphi-\varphi_0)}\right).
\label{U4}
\end{equation}
The corresponding field configuration is also given implicitly:
\begin{equation}
\varphi(t) = \mp \frac13 {\rm arccosh}
\left(\frac{\left(h^2(t)+\frac{\dot{h}(t)}{3}\right)}{\sqrt{A}} +
\sqrt{\frac{\left(h^2(t)+\frac{\dot{h}(t_0)}{3}\right)^2}{A} -
1}\right) + \varphi_0. \label{implicit2}
\end{equation}
Similarly, using the dependence of $\varepsilon$ on $a$ one can
reconstruct the tachyon potential:
\begin{equation}
V(T) = \sqrt{A} = const. \label{V4}\end{equation} It is easy to
see \cite{FKS} that the tachyon model with a constant potential is
exactly equivalent to the Chaplygin gas model. Indeed, in the case
of a constant tachyon potential the relation between the tachyon
energy density (\ref{energy}) and the pressure (\ref{pressure}) is
just that of the Chaplygin gas (\ref{model4}), where $p\
\varepsilon = -V^2(T) = -A$. The Chaplygin gas cosmological model
was introduced in \cite{we} and further developed in
\cite{Fabris,Bilic,Bento,we1} and many other papers. Comparison
with observational data has also been extensively performed
\cite{Chap-obs}.
\end{enumerate}

\section{Transgressing the boundaries}
\label{sect4} We now take a step back and consider the problem of
finding a tachyonic field theory admitting the same cosmic
evolution as the one produced by perfect fluid with equation of
state $p = k\varepsilon$, where now $k > 0$. As we have stated, it
is impossible to reproduce this dynamics using Sen's tachyonic
action (\ref{tachyon}). One way out is to introduce a new field
theory based on a Born-Infeld type action with Lagrangian
\begin{equation}
L = W(T)\sqrt{\dot{T}^2-1}. \label{Lagn}
\end{equation}
In this new field theory the energy and pressure are given by
\begin{equation}
\varepsilon = \frac{W(T)}{\sqrt{\dot{T}^2-1}}, \label{enn}
\end{equation}
and
\begin{equation}
p = W(T)\sqrt{\dot{T}^2-1}. \label{prn}
\end{equation}
The pressure (if well defined) is now positive. On the other side,
the equation of motion for this field has exactly the same form
(\ref{tac-eq}) that was derived by Sen's action.

Following the procedure described in Section 2, now applied to the
Lagrangian (\ref{Lagn}), one gets the following potential
corresponding to the equation of state $p = k\varepsilon$ (with $k
> 0)$:
\begin{equation}
W(T) =  \frac49 \frac{\sqrt{k}}{(1+k)}\frac{1}{T^2}. \label{potn}
\end{equation}
The exact solution of the field equations that reproduces the
dynamics of the perfect fluid is
\begin{equation}
T(t) = \sqrt{1+k}\ t \label{exactn}
\end{equation}
(we restrict our attention to the region of the phase space where
$T \geq 0$ and $\dot{T} \geq 1$). There are two other obvious
solutions for this model, corresponding to other choices of the
initial conditions; they also give rise to linearly growing
fields: $T(t) = t$ and $T(t) = \sqrt{1+ \frac{1}{k}}\ t$.

We would like to point out an interesting fact: the Lagrangian
(\ref{Lagn}) which, together with the explicit form (\ref{potn})
of the potential, describes the field theory corresponding to a
positive value of $k$, is actually the same as Sen's Lagrangian
(\ref{tachyon}) with  potential (\ref{V1}) itself considered for
positive $k$. Indeed, it is true on the one hand that in this case
the potential (\ref{V1}) becomes imaginary. However, this can be
compensated by considering the kinetic term in the region
$1-\dot{T}^2 < 0$ so that the action as whole remains real.  It
can be re-interpreted as the product of two real terms:
\begin{equation}
L =- V(T)\sqrt{1-\dot{T}^2} = (\sqrt{-1})^2 V(T)
\sqrt{1-\dot{T}^2} = W(T) \sqrt{\dot{T}^2-1}.
\end{equation}
This model is introduced here as a pedagogical introduction to the
model with the potential (\ref{V2}), which we discuss in detail in
the following section. The properties of the present model can be
recovered in the limit $\Lambda \to 0$ and we will not further
comment on it.

\section{Dynamics of the toy tachyonic model}
\begin{figure}[h]
\epsfxsize 7.45cm \epsfbox{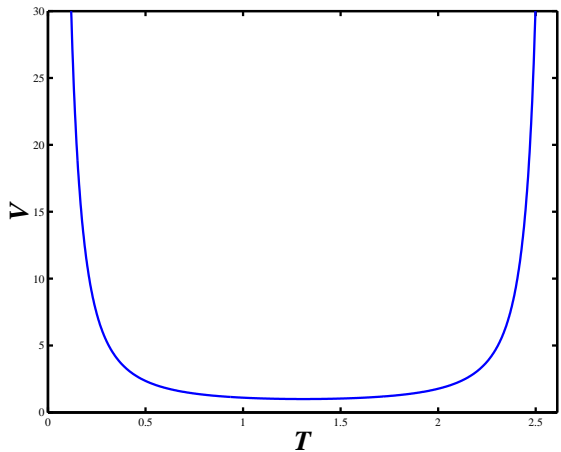} \hfill \epsfxsize 7.7cm
\epsfbox{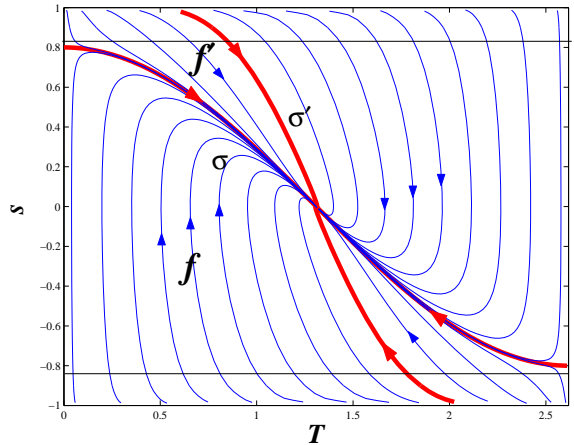} \caption{Potential and phase portrait for
$k<0$ (here $k=-0.36$). The horizontal lines $s = \pm \sqrt{2/3}$
separate the central region where the expansion of the universe is
accelerated from the two external regions where it decelerates.}
\label{Fig1}
\end{figure}

\begin{figure}[h]
\epsfxsize 7.6cm \epsfbox{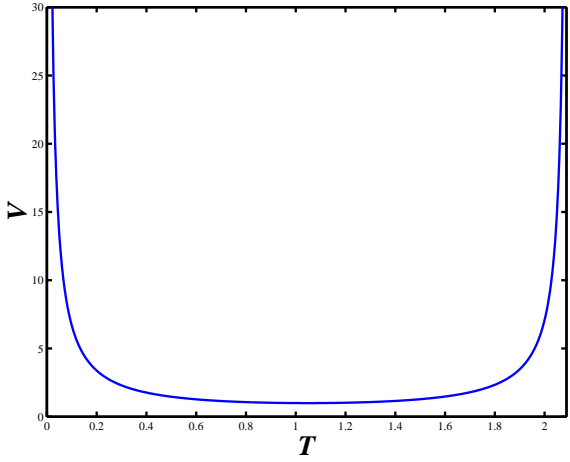} \hfill \epsfxsize 7.7cm
\epsfbox{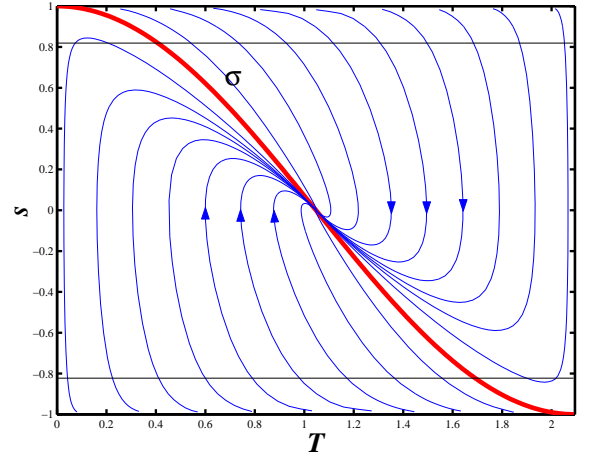} \caption{Potential and phase portrait for
$k=0$. Since $k>-1/3$ there are trajectories which undergo two
epochs of deceleration and two epochs of acceleration.}
\label{Fig2}
\end{figure}

We now provide the analysis of the dynamics of the tachyonic model
based on the potential (\ref{V2}). In this case eq. (\ref{tac-eq})
is equivalent to the following system of two first-order
differential equations \
\begin{eqnarray}
&& \dot{T} = s, \label{eq1}\\
&& \dot{s} = -3\sqrt{V}(1-s^2)^{3/4}s -(1-s^2)\frac{V_{,T}}{V},
\label{eq2}
\end{eqnarray}
where using the Friedmann equation (\ref{Friedmann1}) we have
expressed the Hubble variable $h$ as a function of the variables
$T$ and $s$. The model has the following two exact solutions (we
take $T_0 = 0$ without loss of generality):
\begin{eqnarray}
&& T_1(t) = \frac{2}{3\sqrt{\Lambda(1+k)}}\arctan \sinh
\frac{3(1+k) \sqrt{\Lambda}t}{2}, \ \ 0<t<\infty,\label{solution}
\\
&& T_2(t) = \frac{2}{3\sqrt{\Lambda(1+k)}}\left(\pi - \arctan
\sinh \frac{3(1+k)\sqrt{\Lambda}t}{2}\right) \ \ 0<t<\infty.
\label{solution1}
\end{eqnarray}
By inserting (\ref{V2}) into (\ref{eq2})    and by eliminating the
time we obtain an equation for the phase space trajectories
$s=s(T)$:
\begin{eqnarray}
&&\frac{ds}{dT} = -\frac{3(1-s^2)\sqrt{\Lambda}} {\sin
\frac{3\sqrt{\Lambda(1+k)}T}{2}} \left(\frac{1-(k+1)\cos^2
\frac{3\sqrt{\Lambda(1+k)}T}{2}}{1-s^2}\right)^{1/4} +
\nonumber \\
&&-\frac{3\sqrt{\Lambda(1+k)}}{2} \frac{1-s^2}{s}\cot
\left(\frac{3\sqrt{\Lambda(1+k)}T}{2}\right)
\frac{(k+1)\cos^2\frac{3\sqrt{\Lambda(1+k)}T}{2} + (k-1)}{1 -
(k+1)\cos^2 \frac{3\sqrt{\Lambda(1+k)}T}{2}}. \label{eq5}
\end{eqnarray}
In the phase plane $(T,s)$ the solutions (\ref{solution})
(\ref{solution1}) correspond to arcs of the curve $\sigma$ (see
Figs. \ref{Fig1}, \ref{Fig2} and \ref{Fig3}):
\begin{equation}
s = \sqrt{1+k} \, \cos \frac{3\sqrt{\Lambda(1+k)}\,T}{2}
\label{solution2}
\end{equation}
The behavior of the cosmological radius for both solutions
(\ref{solution}) and (\ref{solution1}) is the following:
\begin{equation}
a(t) = a_0 \left(\sinh
\frac{3\sqrt{\Lambda}(1+k)t}{2}\right)^{\frac{2}{3(1+k)}}.
\label{cosm-rad}
\end{equation}
As expected, we get back the cosmological evolution
(\ref{Hubblemod2}) determined by the equation of state
(\ref{model21}) which was the starting point for constructing the
potential (\ref{V2}).  To study the cosmological evolutions
corresponding to all possible initial conditions we need to
distinguish two different cases. When the parameter $k\leq 0$
there are no surprises and the associated cosmology is essentially
driven by that of the exact solutions while when $k$ is positive
the model seems at first to be ill-defined. We will have again to
go beyond the model itself and "transgress the boundaries" to see
what its possible meaning can be.

\vskip10pt

In the first case when $-1 < k \leq 0$ the potential (\ref{V2}) is
well-defined for
\begin{equation}
0 < T < \frac{2\pi}{3\sqrt{\Lambda(1+k)}}; \label{range}
\end{equation}
while the dynamics guarantees that \begin{equation} -1 < s < 1.
\label{range1}
\end{equation}
The system has only one critical point, namely
\begin{equation}
s_0 = 0,\ T_0 = \frac{\pi}{3\sqrt{\Lambda(1+k)}}.
\label{special}
\end{equation}
The eigenvalues of the linearized system in the neighborhood of this point are
\begin{equation}
\lambda_{1,2} = -\frac32 \sqrt{\Lambda}(1 \pm k).
\label{eigen}
\end{equation}
Both of them are real and negative. Thus, this special point is an
attractive node. It corresponds to a de Sitter expansion with a
Hubble parameter
\begin{equation}
H_0 = \sqrt{\Lambda}.
\label{Hubble}
\end{equation}
The set of integral curves of (\ref{eq5}) is symmetric under
reflection with respect to the critical point (\ref{special}): any
given integral curve and its node (n)-reflected one describe the
same cosmological evolution. The  curve $\sigma$ corresponds to
the eigenvalue $\lambda_1 = -\frac32 \sqrt{\Lambda} (1+k)$, whose
absolute value is the smallest of the two. It acts as a separatrix
for the integral curves. Almost all curves approaching the node
(\ref{special}) end up there with the same tangent as $\sigma$.
The only exception is a second separatrix $\sigma'$ which
corresponds to the eigenvalue $\lambda_2 = -\frac32 \sqrt{\Lambda}
(1-k)$. The curve $\sigma'$ separates the bundle $f'$ of the
curves which do not intersect the axis $s = 0$ from the bundle $f$
of those which intersect it (see Fig. 1). The boundary of the
rectangle defined by Eqs. (\ref{range}) and (\ref{range1})
describes a cosmological singularity. Indeed, the scalar curvature
for a flat Friedmann universe is
\begin{equation}
R = 6(\dot{h} + 2 h^2).
\label{curvature}
\end{equation}
Since from Eqs. (\ref{Friedmann1}), (\ref{en-cons}),
(\ref{energy}) and (\ref{pressure}) one has that
\begin{equation}
\dot{h} = -\frac32 h^2 s^2 \label{hdot2}
\end{equation}
by substituting into (\ref{curvature}) it follows that
\begin{equation}
R = 3 h^2(4-3s^2) = \frac{3V(T)(4-3s^2)}{\sqrt{1-s^2}}.
\label{curvature1}
\end{equation}
Thus the scalar curvature $R$ tends to infinity when approaching
the boundary of the rectangle (with the exception of the corners,
that will be treated separately).

All integral curves end up in the node; let us see how they behave
close to the boundary and begin with a right neighborhood of $T =
0$. There equation (\ref{eq5})  takes the form
\begin{equation}
\frac{ds}{dT} \approx \frac{2(1-s^2)}{s\,T}\left(1 -
\frac{(-k)^{1/4}s}{\sqrt{1+k}(1-
s^2)^{1/4}}\right)=\frac{F(s)}{T}. \label{eq-as}
\end{equation}
If $s < \sqrt{1+k}$, then $\dot{s} \rightarrow +\infty$ as $T
\rightarrow 0$. Therefore, the integral curves at $s < \sqrt{1+k}$
which get close to the $T = 0$ axis rise almost vertically,
climbing leftwards for $s < 0$ and rightwards for $s > 0$ until
they get close to $\sigma$ whenceforth they reach a maximum and
thereafter approach the de Sitter node (\ref{special}) (see Fig.
1). These are the curves of bundle $f$.  If $\sqrt{1+k} < s < 1$,
then $\dot{s} \rightarrow -\infty$ as $T \rightarrow 0$.
Therefore, the corresponding curves at $s > \sqrt{1+k}$ which get
close to the $T = 0$ axis drop almost vertically until they get
close to $\sigma$ whenceforth they also approach the de Sitter
node. These are the curves of bundle $f^\prime$. On the separatrix
$\sigma$ one attains the point $(0, \sqrt{1+k})$ where $\dot{s} =
0$. Symmetric considerations apply to the n-reflected curves, i.e.
those which lie to the right of the separatrix $\sigma'$.

But where do all the curves originate from? We first show that
apart from $\sigma$, none of them can touch any point of the
$s$-axis. Indeed, let us consider a point $(T,s)$ close to the
$s$-axis. Eq. (\ref{eq-as}) can be integrated backwards to give
\begin{equation}
T(s) = T_0(s_0)\exp \int_{s_0}^s \frac{dx}{F(x)}. \label{eq-as1}
\end{equation}
This equation shows that, if $F(s_0)\not=0$, it is impossible to
realize the condition $T_0(s_0) = 0$ and therefore touch the $s=0$
axis on a given trajectory.  The roots of the equation $F(s_0) =
0$ are $s = \sqrt{1+k}, \pm 1$.  A closely similar reasoning
excludes the point $(0,1)$. The point $(0,-1)$ can also be
excluded since we should have $\dot{T} \geq 0$ in a neighborhood
of such point; but $s$ is negative and this contradicts the
equation $\dot{T} = s$. We are therefore left with the point
$(0,\sqrt{1+k})$, where the exact solution $\sigma$ originates.

Let us examine now the upper boundary of Fig. \ref{Fig1}. In a
small neighborhood of the point $(T_*,1)$ (where $T_*$ is in the
domain (\ref{range})) Eq. (\ref{eq5}) can be replaced by the
following approximate equation:
\begin{equation}
\frac{d {s}}{dT} = -\frac{3(2(1-{s}))^{3/4}\sqrt{\Lambda}\left(1 -
(1+k)\cos^2\frac{3\sqrt{\Lambda(1+k)}T_*}{2}\right)^{1/4}}{\sin
\frac{3\sqrt{\Lambda(1+k)}T_*}{2}}. \label{eq-as4}
\end{equation}
This equation is not Lipschtzian. The upper integral is the
trivial solution $s = 1$ while the lower integral is
\begin{equation}
s  \rightarrow \left\{\begin{array}{lll}
=1 & {\rm for}& T  < T_* \\
\thickapprox 1-C(T_*)(T-T_*)^4 &{\rm for}& T \geq T_*
\end{array}\right. \label{sol-as3}
\end{equation}
where
\begin{equation}
C(T_*) = \frac{81}{32} \frac{\Lambda^2 \left(1 -
(1+k)\cos^2\frac{3\sqrt{\Lambda(1+k)}T_*}{2}\right)}{\sin^4
\frac{3\sqrt{\Lambda(1+k)}T_*}{2}}. \label{C-def}
\end{equation}
The intermediate solutions stay constant at $s = 1$ for a while
and then leave the $s=1$ line at a value $T_{**}>T_*$. Therefore,
from each point $(T_*,1)$ there originates only one integral curve
behaving as (\ref{sol-as3}). In particular, the separatrix
$\sigma'$ originates at a point $(T_{\sigma'},1)$  (the value of
$T_{\sigma'}$ being unknown).

The condition of cosmic acceleration is
\begin{equation}
\frac{2\ddot a}{a} = -(\varepsilon + 3p) > 0. \label{acc-cond}
\end{equation}
For the tachyon cosmological model, using formulas (\ref{energy})
and (\ref{pressure}) this condition can be reexpressed as
\begin{equation}
s^2 < \frac{2}{3}.
\label{accel1}
\end{equation}
Therefore, if $k \leq -\frac13$ all cosmological evolutions
undergo an initial phase of deceleration followed by an
accelerating one. On the other hand, if $k > -\frac13$ all
evolutions whose originating point $T_*$ is larger than a critical
value (which depends on $k$) have two epochs of deceleration and
two epochs of acceleration. This happens in particular in the
limiting case when $k = 0$, as shown in Fig. \ref{Fig2}. \vskip
10pt

We consider now the case $0 < k < 1$. Eq. (\ref{V2}) shows that
the potential $V(T)$ is well defined only in the interval
$(T_3,T_4)$ where
\begin{equation}
T_{3} = \frac{2}{3\sqrt{(1+k)\Lambda}}\arccos
\frac{1}{\sqrt{1+k}}, \ \ \ \ \  T_{4} =
\frac{2}{3\sqrt{(1+k)\Lambda}} \left(\pi - \arccos
\frac{1}{\sqrt{1+k}}\right). \label{T4}
\end{equation}
Together with the condition $-1<s<1$, Eq. (\ref{T4}) defines the
rectangle where we study the model at first. There are now three
fixed points. One of them is the attractive de Sitter node
(\ref{special}) whereas the other two are saddles corresponding to
the maxima of the potential at coordinates
\begin{eqnarray}
&&T_1 = \frac{2}{3\sqrt{(1+k)\Lambda}}\arccos
\sqrt{\frac{1-k}{1+k}}, \;\;\;\;\; T_2 =
\frac{2}{3\sqrt{(1+k)\Lambda}}\left(\pi - \arccos
\sqrt{\frac{1-k}{1+k}}\right). \label{saddle}
\end{eqnarray}
(s=0). They give rise to an unstable de Sitter regime with Hubble
parameter
\begin{equation}
H_1 = \sqrt{\frac{(1+k)\Lambda}{2 \sqrt{k}}}> H_0. \label{Hubble1}
\end{equation}

We first analyze the behavior of the trajectories in the vicinity
of the line $T = T_3$ and set $T = T_3 + \tilde{T}$, with
$\tilde{T}$ small and positive and $s \neq \pm 1$. With these
conditions the model is described by the approximate equation
\begin{equation}
\dot{s} \approx -\frac{1-s^2}{2\tilde{T}}, \label{corner}
\end{equation}
which implies that the trajectories passing close to the boundary
$T = T_3$ drop steeply down without crossing it. The "physical"
reason for this behavior is the vanishing of the potential at
$T=T_3$ (to our knowledge this is a novel feature of our model);
indeed, the structure of the tachyonic action implies that the
"force" is proportional to the logarithmic derivative of the
potential and this is infinite at $T=T_3$. The impossibility of
crossing $T=T_3$ remains true also without coupling the tachyon to
gravity. On the other hand, in contrast with the situation
encountered before, the geometry is regular at $T=T_3$:  the
vertical boundaries of the rectangle are not curvature
singularities, because the potential does not diverge there.
Actually, the curvature scalar $R$ vanishes there because of the
vanishing of the potential.
\begin{figure}[h]
\epsfxsize 10cm \centerline{\epsfbox{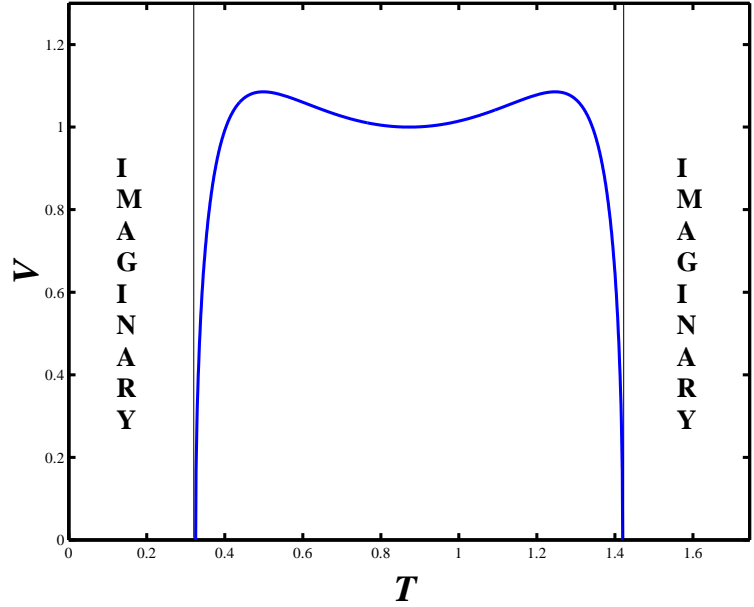}}
\caption{Potential for $k>0$ (here $k=0.44$).} \label{Fig3a}
\end{figure}
\begin{figure}[h]
\epsfxsize 15cm \centerline{\epsfbox{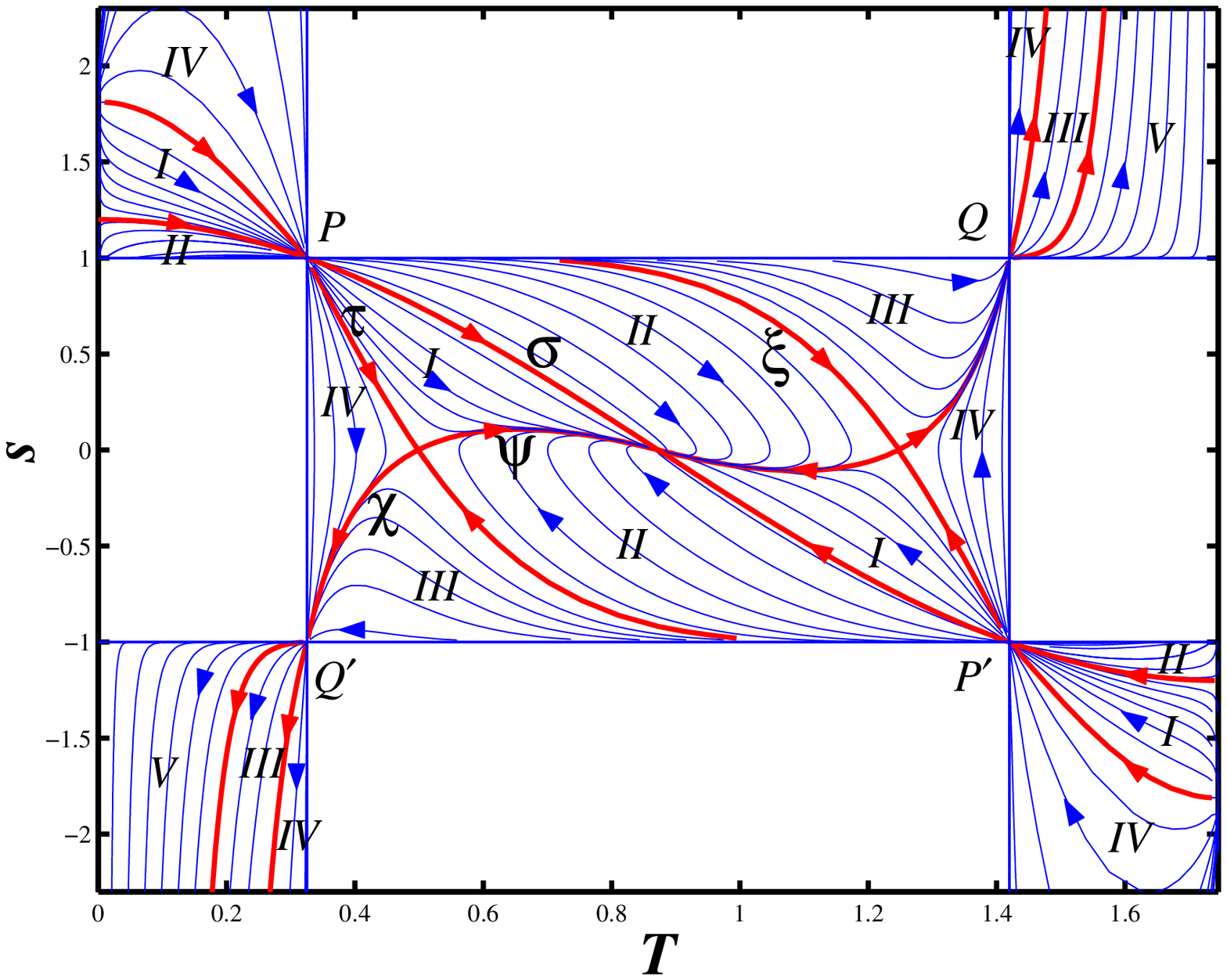}}
\caption{Phase portrait evolution for $k>0$ ($k=0.44$).}
\label{Fig3}
\end{figure}
Instead, the horizontal sides are still singular. The situation is
exactly as before and there is one integral curve which originates
at $(T_*,1)$ whose behavior is again given by (\ref{sol-as3}),
(\ref{C-def}). Now, however, due to the positivity of the
parameter $k$, the nonnegative function $C_k(T_*)$ vanishes at
$T=T_{3}$ and at $T=T_{4}$, is maximal at $T=T_1$  and at $T=T_2$,
and has a positive local minimum at $T = T_0$.

Now consider the behavior of the trajectories in the upper left
vertex of the rectangle $P=(T_3,1)$. Setting $s = 1 - \tilde{s}$
and $T = T_3 + \tilde{T}$, with $\tilde T$ and $\tilde s$ small
but not zero, from Eq. (\ref{eq5}) we get the approximate equation
\begin{equation}
\frac{d\tilde{s}}{d\tilde{T}} - \frac{\tilde{s}}{\tilde{T}} = A
\,\tilde{s}^{3/4}\tilde{T}^{1/4}, \label{corner-eq}
\end{equation}
where $ A = 3^{5/4}2^{3/4}\Lambda^{5/8}(1+k)^{5/8}k^{-3/8}. $ The
general solution is
\begin{equation}
\tilde{s} = \frac{1}{256}(A\tilde{T} + B)^4 \tilde{T},
\label{corner-sol}
\end{equation}
where $B$ is an arbitrary constant. When $B \neq 0$ the leading
behavior is
\begin{equation}
\tilde{s} = D\tilde{T},\,\,D>0. \label{corner-sol1}
\end{equation}
This means that there are trajectories which gush out from the
point $P$ in all possible directions, except of course vertically
(see Fig. 4). If $B = 0$ Eq. (\ref{corner-sol}) becomes
\begin{equation}
\tilde{s} = \frac{A^4}{256}\tilde{T}^5.
\label{corner-sol2}
\end{equation}
This equation describes the leading behavior of a curve $\rho$
which separates the trajectories of type (\ref{corner-sol1}) from
those originating at the points $(T_*,1)$. We point out  that $P$
is a cosmological singularity for the curve $\rho$ (which
therefore originates at $P$), whereas it is regular for the curves
(\ref{corner-sol1}). Indeed, we see from Eq. (\ref{Friedmann1})
that when $\tilde{T} \rightarrow 0\ \  h^2$ behaves as
$1/\tilde{T}^2$ along $\rho$, while it is finite along
trajectories (\ref{corner-sol1}).

Now consider  the upper right vertex $Q = (T_4,1)$. By setting as
before $s = 1 - \tilde{s}$ and $T = T_4 - \tilde{T}$ we get the
approximate equation
\begin{equation}
\frac{d\tilde{s}}{d\tilde{T}} - \frac{\tilde{s}}{\tilde{T}} = -A
\tilde{s}^{3/4}\tilde{T}^{1/4},
\label{corner1-eq}
\end{equation}
where $A$ is the same as above. The general solution is
\begin{equation}
\tilde{s} = \frac{1}{256}(-A\tilde{T} + B)^4 \tilde{T},
\label{corner1-sol}
\end{equation}
where $B$ is an arbitrary positive constant. Therefore the
trajectories enter point $Q$ along all possible directions except
vertically and horizontally. We now classify the behavior of the
trajectories in the interior of the rectangle. First note that
there are five distinguished trajectories (separatrices): $\sigma$
which connects  $P$ with the node;  $\tau$ which connects $P$ with
the saddle $(T_1,0)$; $\xi$, which is the curve which enters the
saddle $(T_2,0)$ from above (We are unable to say whether $\xi$
originates at some point $(T_*,1)$ or it belongs to the family
(\ref{corner-sol1}), or it coincides with the curve $\rho$ defined
by equation (\ref{corner-sol2}): for this reason we have not tried
to draw curve $\rho$ in Fig \ref{Fig3}); $\psi$, which originates
from the saddle $(T_1,0)$ and enters the node with tangent defined
by the eigenvalue $\lambda_2 = -\frac32 \sqrt{\Lambda}(1-k)$ (see
(\ref{eigen})); finally, $\chi$ which connects $(T_2,0)$ with $Q$
(see Fig. \ref{Fig3}). Each of these separatrices has its own
n-reflected counterpart (denoted by the same symbol).

Corresponding to the separatrices $\sigma, \tau, \xi, \psi$ and
$\chi$ one can distinguish four bundles of qualitatively different
trajectories. 1) The bundle $f^I$ of trajectories limited by
$\tau, \psi$ and $\sigma$: they originate from $P$ and enter the
node along $\psi$ and from above: 2) the bundle $f^{II}$ of
trajectories limited by $\sigma, \xi$ and $\psi$ : they enter the
node along $\psi$ from below; 3) the bundle $f^{III}$ of
trajectories limited by $\xi,\chi$ and the horizontal line $s =
1$; they stream into $Q$; 4) the bundle $f^{IV}$ of the curves
which are limited by $\tau, \chi$ and the vertical line $T = T_3$:
they gush out of $P$ and stream into $Q'$.

Now we have to face a problem that we have not mentioned so far.
The question is the following: it takes a finite proper time for
the fields (and the universe) to get from anywhere to the corner
$Q$ or $Q'$ on a trajectory of bundle $f^{III}$ or of bundle
$f^{IV}$. But these corners are not critical points of the
dynamical system and, furthermore, the universe does not
experience any singularity by getting there along these
trajectories. Similar remarks apply to the past of the
trajectories originating from $P$ and $P'$. If the model could not
be extended to follow these trajectories outside the rectangle
where it has been originally defined it would be useless. However,
we now show that this extension is actually possible. Indeed, one
can see by inspection that the field equations are well defined
also inside the four semi-infinite strips defined by the following
inequalities: $0 < T < T_3$, with $s > 1$ or $s < -1$; $T_4 < T <
\frac{2\pi}{3\sqrt{\Lambda(1+k)}}$ with $s > 1$ or $s < -1$. Then,
following the strategy sketched in Section \ref{sect4}, we
introduce a "new" Lagrangian:
\begin{equation}
L = W(T)\sqrt{\dot{T}^2-1}, \label{Lagrange1}
\end{equation}
where the "new" potential is given by
\begin{equation}
W(T) = \Lambda \frac{\sqrt{(k+1)\cos^2
\frac{3\sqrt{\Lambda(1+k)}T}{2}-1}}
{\sin^2\frac{3\sqrt{\Lambda(1+k)}T}{2}}. \label{potential1}
\end{equation}
The "new" Lagrangian comes itself out of Sen's action (Eqs.
(\ref{tachyon}),(\ref{V2})) by the previous trick:
\begin{eqnarray}
&& V(T) \longrightarrow W(T) = i V(T) \\\label{ima} &&
\sqrt{1-\dot{T}^2} \longrightarrow i \sqrt{1-\dot{T}^2} =
\sqrt{\dot{T}^2-1}\label{ima1}
\end{eqnarray}
i.e. both the "old" kinetic term and the potential become
imaginary but their product remains real. It follows from the
dynamics (\ref{eq5}) that the expressions under the square root in
the potential and the kinetic term change sign simultaneously. In
other words, in the phase diagram no trajectory can cross any side
of the rectangle. Thus, all the quantities characterizing the
model keep real during the dynamical evolution. White regions in
the phase diagram, where the Lagrangian and other quantities would
become imaginary, are forbidden!
 The product
(\ref{Lagrange1}), which amounts exactly to the "old" Lagrangian,
can be interpreted in terms of the "new" kinetic term and
potential that are both real. This makes it clear why the "new"
Lagrangian gives rise to field equations in the above four strips,
which are the same as in the interior of the rectangle. At first
glance one might have the impression that there is a freedom of
choice of sign for the new Lagrangian (\ref{Lagrange1}), or, in
other words, that one may choose opposite signs in Eqs.
(\ref{ima}), (\ref{ima1}). However, this is not so: the choice of
sign in Eq. (\ref{Lagrange1}) is determined by the requirement of
continuity of the Einstein (Friedmann) equations when passing from
the rectangle to the stripes.
 As anticipated in Sec.
\ref{sect4}, both the energy density and the pressure  are
positive in the considered strips.

However, there is an important difference between the present
situation and the one described in Sec. \ref{sect4}. Here we are
dealing with one single model ($k$ is fixed). One is compelled to
make the extension of the model and, contrary to the $\Lambda=0$
case, the two different "phases" (i.e.  negative and positive
pressure) are found within the same model, at different stages of
the cosmic evolution, one phase in the rectangle, the other in the
strips. In the following we give the precise mathematical meaning
of this extension. This opens the way to the study of a new class
of tachyon field theories.

We start by describing the behavior of the trajectories in the
lower left strip (see Fig. \ref{Fig3}). Since $\dot{T} < -1$, the
evolution along any given trajectory will lead us in a finite
amount of time to either hit the vertical line $T = 0$ or to
approach a vertical asymptote $T = T_B$, with $0 \leq T_B < T_3$.
The first alternative does not take place, whereas all values of
$T_B$ in the indicated range are allowed, with the exception of
$T_B = 0$. Indeed, in the vicinity of $T = 0$ Eq. (\ref{eq5})
takes the following form
\begin{equation}
\frac{ds}{dT} = \frac{2(s^2-1)^{3/4}k^{1/4}}{\sqrt{1+k}\,T} -
\frac{2(s^2-1)}{sT}. \label{closetosing}
\end{equation}
Now, assume there is an $s_0 < -1$ such that $\lim_{T \rightarrow
0} s(T) = s_0$. Then, the analogue of Eq. (\ref{eq-as1}) gives a
contradiction: the function $F(s)$ cannot vanish since the r.h.s.
of Eq. (\ref{closetosing}) is positive.

The leading behavior of the solutions of Eq. (\ref{closetosing})
for $T \rightarrow 0$, $s \rightarrow -\infty$, is given by
\begin{equation}
|s|^{1/2} \approx \frac{1}{\frac{k^{1/4}}{\sqrt{1+k}}\ln
\left(\frac{T}{T_B}\right)} \label{TB1}
\end{equation}
where $T_B$ is an arbitrary positive constant. It follows that the
line $T = 0$ is not an asymptote for any trajectory but of course
there are trajectories whose asymptote is as close as we like to
the line $T = 0$.

Therefore, the trajectories inside the considered strip  can be
parameterized by the value $T_B$ of the coordinate of their
vertical asymptote. We now discuss how they behave in the
neighborhood of $s = -1$. As before we can divide the trajectories
inside the strip into three families according to their leading
behavior
\begin{eqnarray}
&& s \approx - 1 + E(T - T_3),\  E > 0, \label{fam1}
\\
&& s \approx - 1 + \frac{A^4}{256}(T - T_3)^5, \label{fam2}
\\
&& s \approx - 1 + C(T_*)(T - T_*)^4,\ 0 < T_* < T_3 \label{fam3}
\end{eqnarray}
Trajectories of type (\ref{fam1}) fan out from point $Q'$ into the
strip at all possible angles. Trajectories of type (\ref{fam3})
can be thought of as originating from $(T_*,-1)$. Note that, just
as before, the function $C(T_*)$ approaches (minus) infinity as
$T_* \rightarrow 0$. It approaches zero as $T_* \rightarrow T_3$.
Curve (\ref{fam2}) separates the families (\ref{fam1}) and
(\ref{fam3}). The coordinate $T_B$ of the asymptote is an
increasing function of $T_*$. As $T_* \rightarrow T_3$ it attains
a value characterizing the asymptote of curve (\ref{fam2}), beyond
which it becomes an increasing function of $E$.

Let us go back to physics and consider the behavior of the
cosmological radius $a(t)$ when the tachyon field $T(t)$ tends to
$T_B$ along the solution of the field equations. The Friedmann
equation implies that $h^2  \rightarrow 0$ and that $\dot{h} \to
-\infty$ (see Eq. (\ref{hdot2})). Therefore, the scalar curvature
(\ref{curvature}) diverges and  the universe  reaches  a
cosmological singularity in a finite time.

This is an unusual type of singularity which we call {\it big
brake}. Indeed, since  ${\ddot{a}}/{a} = \dot{h} + h^2$, in a big
brake we have that
\begin{eqnarray}
&&\ddot{a} \rightarrow -\infty, \nonumber \\
&&\dot{a} \rightarrow 0, \nonumber \\
&&a \rightarrow a_B < \infty\label{Br1}
\end{eqnarray}

In other words, the evolution of the universe comes to a
screeching halt in a finite amount of time and its ultimate scale
depends on the final value $T_B$ of the tachyon field.

We now turn to the behavior of the trajectories in the upper left
strip. In the vicinity of $T = 0$ the equation for the
trajectories takes the form (\ref{closetosing}). The coefficient
of ${T}^{-1}$ vanishes at the values
\begin{equation}
s = 1,\ \sqrt{1+k}, \ \sqrt{1+ \frac{1}{k}}. \label{points}
\end{equation}
As explained earlier it is only at these values of $s$ that the
trajectories can leave the line $T = 0$. In addition, an analysis
similar to the one performed earlier shows that the only
trajectory starting at $(0,1)$ is the line $s = 1$, and that the
only trajectory starting at $(0,\sqrt{1+k})$ is the curve
$\sigma$. All other trajectories start from
$\left(0,\sqrt{1+\frac{1}{k}}\right)$. Eq. (\ref{closetosing})
shows that as $T \rightarrow 0$ the derivative $\frac{ds}{dT}$
approaches $-\infty$ for $\sqrt{1+k} < s < \sqrt{\frac{1+k}{k}}$
whereas it approaches $+\infty$ for $1 < s < \sqrt{1+k}$ and $s >
\sqrt{\frac{1+k}{k}}$ (see Fig. \ref{Fig3}). To study the behavior
in the neighborhood of $\left(0,\sqrt{\frac{1+k}{k}}\right)$ we
set $s = \sqrt{\frac{1+k}{k}} + \tilde{s}$ with $\tilde{s}$ small.
In the approximate equation for $\tilde{s}(T)$ it is necessary,
besides the leading term , to keep the term proportional to $T$.
Thus we get
\begin{equation}
\frac{d\tilde{s}}{dT} = \frac{2(1-k)}{1+k}\frac{\tilde{s}}{T} -
\frac{9\Lambda}{8}\sqrt{\frac{1+k}{k}}\left(\frac{k+3}{k}\right)T.
\label{point1}
\end{equation}
The general solution of this equation is
\begin{equation}
\tilde{s} = \left(D -
\frac{9\Lambda(1+k)^{3/2}(k+3)}{32k^{5/2}}T^{\frac{4k}{1+k}}\right)T^{\frac{2(1-k)}{1+k}},
\label{point2}
\end{equation}
where $D$ is an arbitrary constant. If $D\not = 0$ we have the
leading behavior
\begin{equation}
{s} \approx \sqrt{1+\frac{1}{k}} + D T^{\frac{2(1-k)}{1+k}}
\label{point3}
\end{equation}
Therefore, if $k < \frac13$ the trajectories start from $\left(0,
\sqrt{1+\frac{1}{k}}\right)$ with horizontal tangent, whereas they
start with vertical tangent when $k > \frac13$. If $k = \frac13$
they are born with any possible tangent. When $ D = 0$ we must go
to the next order:
\begin{equation}
{s} \approx \sqrt{1+\frac{1}{k}} -
\frac{9\Lambda(1+k)^{3/2}(k+3)}{32k^{5/2}}T^2. \label{point5}
\end{equation}
This curve acts as a separatrix for the curves having positive and
negative $D$ respectively. Regarding the curves of the strip which
lie below $\sigma$ they depart from $(T_*,1)$ and behave in the
neighborhood of this point as
\begin{equation}
s = 1 - C(T_*)(T-T_*)^4,\ 0 < T_* < T_3, \label{point6}
\end{equation}
where the function $C(T_*)$ is the same as in (\ref{fam3}).
Regarding the behavior of the trajectories in the neighborhood of
$T=T_3$ a simple analysis shows that they stream into $P$ at all
possible angles (except vertically and horizontally). These
properties show that each curve of type (\ref{point3}) with
positive $D$ attains a maximum somewhere between $T = 0$ and $T =
T_3$ whose height is an increasing function of $D$ which tends to
infinity as $D \rightarrow +\infty$.

So far, we have analyzed the behavior of the trajectories in two
distinct regions: the rectangle and the four strips. Now, it has
to be noted that the trajectories in the rectangle which leave $P$
(or $P'$) at all possible angles in the open interval $(0,\pi/2)$
and the trajectories which enter $Q'$ (or $Q$), again at all
possible angles, are incomplete, since the vertices of the
rectangle are not cosmological singularities for these curves. The
same is true for the trajectories in the strips which enter $P$
(or $P'$) and leave $Q$ (or $Q'$). This circumstance, and the fact
that the equation of motion for $T$ is the same in the rectangle
and in the strips, indicates that it must be possible to extend
and complete the above set of trajectories by continuation through
the vertices of the rectangle. Precisely, the trajectories
entering $P$ (or $P'$) from the upper left (or lower right) strip
shall be continued into the trajectories entering into the
rectangle from $P$ (or $P'$). Similar remarks apply to the corners
$Q$ and $Q'$. The uniqueness of this continuation procedure can be
proved by applying the $\sigma$-process method of resolution of
singularities (see, e.g. \cite{Arnold}). This amounts to blowing
up (unfolding) the vertices of the rectangle by transition to
suitable projective coordinates, the use of which removes the
degeneracy of the vector field at these points. We do not give the
mathematical details.

\section{Cosmology}

The rich mathematical structure that we have exhibited in the
previous section gives rise to the possibility that cosmologies
that have very different features coexist within the same model.
In this sense tachyonic models are richer than the "corresponding"
standard scalar field models.

We now  characterize all the evolutions in our model that are
portrayed in Fig. \ref{Fig3}. We start from the trajectories
originating from the cosmological singularity $\left(0, \sqrt{1+
\frac{1}{k}}\right)$ and which are characterized in the
neighborhood of the latter by formula (\ref{point3}) with the
parameter $D$ positive and very large. One such trajectory, as
soon as it leaves the singularity, raises steeply upwards until it
attains some maximum value. Henceforth it turns steeply downwards,
enters the rectangle at $P$ at some small angle $\alpha$ with the
vertical axis $T = T_3$ and moves towards $Q'$. Upon reaching this
point it enters the lower left strip and eventually ends up in a
finite time $t_B$ in a big brake corresponding to some field value
$T_B$ very close to $T_3$ (inf $t_B = 0)$.

As $D$ decreases the height of the maximum decreases accordingly,
$\alpha$ and $t_B$ increase and $T_B$ decreases. Eventually, at
some critical value $D_c$ of the parameter $D$, the trajectory
degenerates inside the rectangle with the separatrices $\tau$ and
$\chi$ entering and, respectively, leaving the saddle point
$(T_1,0)$. The curves for which $D > D_c$ belong to the bundle
$f^{IV}$. For $D < D_c$ we get the bundle $f^I$. These
trajectories correspond to evolutions which are asymptotically  de
Sitter with a Hubble parameter $H_0 = \sqrt{\Lambda}$. The upper
bound of $D$ for the curves of bundle $f^I$ is $D_c$ and
corresponds to the separatrix $\tau$ and $\psi$. This means that a
tiny difference in the initial conditions will end up into
dramatically different evolutions: one universe goes into an
accelerating expansion of the de Sitter type and the other ends
with a big brake. This should not be confused with a chaotic
behavior: the two evolutions are almost indistinguishable for a
very long time and then suddenly diverge from each other. As $D$
approaches $-\infty$ we end up into the separatrix $\sigma$
originating at the cosmological singularity $(0,\sqrt{1+k})$.

Proceeding further we encounter those evolutions which start at
$(T_*,1)$, $T_* < T_3$. In the neighborhood of the starting point
they behave according to Eq. (\ref{point6}), then enter the
rectangle through $P$ passing above the curve $\sigma$ from below.
In the limit $T_* = T_3$ we obtain the trajectory $\rho$. As $T_*$
increases further beyond $T_3$ the curves detach themselves from
the axis $s = 1$ according to the equation (\ref{point6}) now with
$T_3 < T_* < T_4$. At some critical value $T_*^c$ of $T_*, 0 <
T_*^c < T_4$, the trajectory degenerates into the separatrices
$\xi, \psi$ and $\chi$. The curves for which $0 < T_* < T_*^c$
form the family $f^{II}$ and they asymptotically arrive at the de
Sitter node $(T_0,0)$ from below the axis $s = 0$. Those for which
$T_*^c < T_* < T_4$ form the family $f^{III}$; they enter the
upper right strip through $Q$ and eventually end up in a finite
time in a big brake. Finally, the trajectories that detach
themselves from the axis $s = 1$ at $T_* \geq T_4$ bend again
upwards and they form another bundle that we denote by $f^V$. They
also end up in a big brake.

\begin{figure}[h]
\epsfxsize 14cm \centerline{\epsfbox{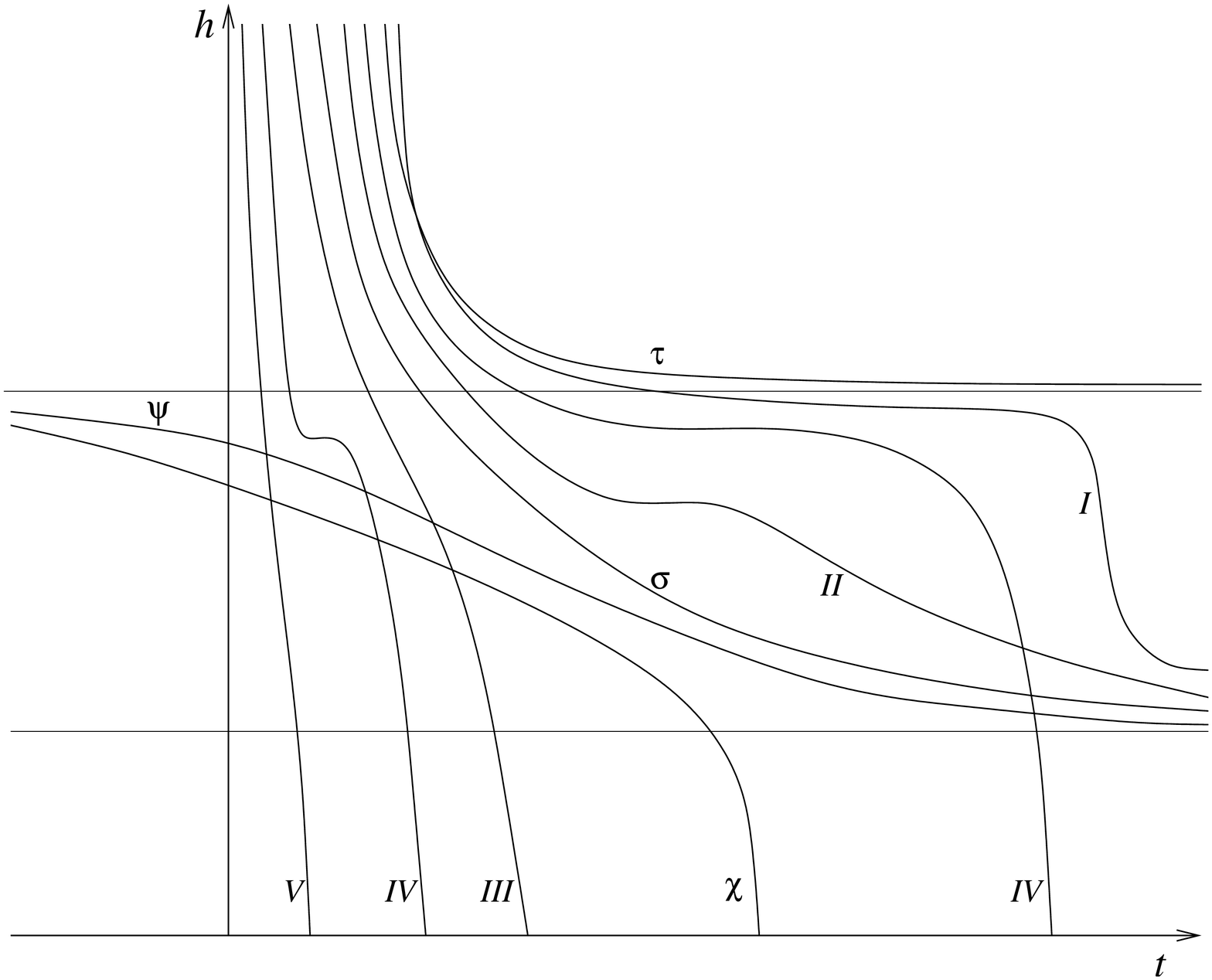}} \caption{Time
evolution of the Hubble parameter $h(t)$.} \label{Fig5}
\end{figure}

The evolutions corresponding to the trajectories of the different
bundles can also be studied in terms of the qualitative behavior
of the Hubble parameter $h(t)$ as a function of time (see Fig.
\ref{Fig5}). In our model, using equations (\ref{Friedmann1}),
 (\ref{V2}) and (\ref{potential1})  we get
\begin{equation}
h(t) =\frac{
\sqrt{\Lambda}}{\sin\left(\frac{3\sqrt{\Lambda(1+k)}}{2}T(t)\right)}
\left|\frac{1-(1+k)\cos^2\left(\frac{3\sqrt{\Lambda(1+k)}}{2}T(t)\right)}{1-s^2(t)}\right|^\frac{1}{4}.
\label{1.1}
\end{equation}

Since $ \dot{h} = -\frac32 h^2 s^2 , \;$ $h(t)$ is positive and
strictly decreasing, except at those values of $t$ for which $s(t)
= 0$.

Close to the initial singularity, the trajectories can be divided
into two classes depending on the singular point from which each
of them starts: the class (A) of trajectories originating at the
point $(0,\sqrt{(1+k)/k})$ and the class (B) of trajectories which
start from the points $(T_* , 1)$. The two classes are separated
by the curve $\sigma$. The leading behavior of $h(t)$ in the
neighborhood of the singularity at $t = 0$ for the trajectories of
class (A) is:
\begin{eqnarray}
&h(t) = \frac{2k}{3(1+k)} \frac 1t - \frac{Dk}{3}
\left(\frac{1+k}{k}\right)^\frac{1-3k}{2(1+k)} t^\frac{1-3k}{1+k}
+ \ldots ,\;\;\;\;\;&{\rm for \ \ }D \not =    0;\label{3.2}\\
&h(t)= \frac{2k}{3(1+k)} \frac 1t - \frac{k^2 - 2k + 9}{32k}
\Lambda t + \ldots, \;\;\;\;\;&{\rm for \ \ }D  =    0
.\label{3.3}
\end{eqnarray}
For trajectories of class (B) the leading behavior is universal
(it does not depend on $k$):
\begin{equation}
h(t)  = \frac{2}{3t} + \ldots \label{4.1}
\end{equation}

To study the behavior of $h(t)$ at later times, one needs to
examine  each bundle $f$ separately, which, in turn, requires
first plotting $h(t)$ for the different separatrices
$\sigma,\tau,\xi,\psi$ and $\chi$.

We have from  Eq. (\ref{cosm-rad})   that $h(t)$ for the
separatrix $\sigma$ is given by
\begin{equation}
h_\sigma(t) =
\sqrt{\Lambda}\coth\left(\frac{3\sqrt{\Lambda}(1+k)t}{2}\right)
\end{equation}
In particular, in the neighborhood of $t = 0$,
\begin{equation}
h_\sigma(t) = \frac{2}{3(1+k)}\frac{1}{t} + \frac{\Lambda(1+k)}{2}
t + \ldots
\end{equation}
to be compared with Eqs. (\ref{3.2}), (\ref{3.3}) and (\ref{4.1}).
Furthermore,   $h_\sigma(t)$    is strictly decreasing and, as
$t\to \infty$, it approaches a stable de Sitter expansion with
Hubble parameter $\sqrt\Lambda$.

Regarding the unstable cosmological evolution   $h_\tau(t)$, for
small $t$ we have
\begin{equation}
h_\tau(t) =\frac{2k}{3(1+k)}\frac{1}{t} - \frac{D_c k}{3}
\left(\frac{1+k}{k}\right)^\frac{1-3k}{2(1+k)}t^\frac{1-3k}{1+k} +
\ldots
\end{equation}
and $ \lim_{t\to \infty} h_\tau(t) = H_1$ (see Fig. \ref{Fig5} and
Eq. (\ref{Hubble1})). Qualitatively, $h_\tau(t)$ behaves similarly
to $h_\sigma(t)$       but the  asymptotic value of the Hubble
parameter is higher. As for the separatrix $\psi$, the
corresponding cosmological evolution is unstable and non singular:
the Hubble parameter decreases steadily from an unstable de Sitter
regime $h_{\psi}({-\infty})= H_1$ at large negative times to a
stable one $h_{\psi}(\infty)= \sqrt{\Lambda}$ at large positive
times. $h_\xi(t)$  and $h_\tau(t)$ have qualitatively similar
behaviors at large times while they are different at small times.
Finally, $h_\chi(t)$ decreases steadily from an unstable
asymptotic de Sitter value $h_{\chi}({-\infty})= H_1$ at large
negative times to the final big brake singularity $h_\chi(t_B)=0$,
with $\dot h_\chi(t_B) = -\infty$, for some suitable $t_B$.

For a trajectory of bundle  $f^I$ which lies close to the
sepratrices   $\tau$  and $\psi$, $h(t)$ behaves according to
(\ref{3.2}) close to the initial singularity,
 with   $D<D_c$, $(D_c-D)$
small. Then $h(t)$ decreases steadily and remains close to
$h_\tau(t)$ for a very long time during which $s$ (and hence $\dot
h $) gets close to zero. Therefore, in this regime the evolution
simulates an asymptotic de Sitter expansion with Hubble parameter
close to the value $H_1$. Eventually, however, $s$ starts
increasing again and the graph of $h(t)$ bends downwards away from
$h_\tau(t)$ and approaches asymptotically the stable de Sitter
regime with Hubble parameter $\sqrt{\Lambda}$. Instead, for a
trajectory of bundle $f^I$ which lies very close $\sigma$ the
behavior $(\ref{3.2})$ of $h(t)$ at small times is characterized
by a value of the constant $D$ which is negative and very large;
the graph of $h(t)$ parts only slightly from the graph of
$h_\sigma(t)$ and the asymptotic value $\sqrt{\Lambda}$ is
approached much earlier. Other elements of $f^I$ display behaviors
which are intermediate between those described above. The time
dependence of $h$ for the trajectories of bundle $f^{II}$ is
qualitatively similar to the one relative to the curves of $f^I$,
the differences being the following: the small time behavior is
given by (\ref{4.1}), and for each trajectory of $f^{II}$ there is
a value $t_0$ of $t$ (which depends on the particular trajectory)
for which $s(t_0)= 0$ so that $\dot{h}(t_0)= 0$.

Now consider a curve of bundle $f^{III}$    which lies very close
to the separatrices $\xi$ and $\chi$. For such a curve $h(t)$
remains very close to $h_\xi(t)$         for a very long time,
$s(t)$ decreases getting close to zero and the evolution simulates
again an asymptotic de Sitter expansion with Hubble parameter.
However, eventually $s(t)$ starts increasing indefinitely and in a
finite (though long) time the cosmology end up in a big brake.
Moving to curves that are farther and farther away from $\xi$ and
$\chi$, the value $T _*$   of $T$ at the initial singularity moves
to the right towards the value $2\pi/(3\sqrt{\Lambda(1+k)})$,
$h(t)$ decreases more and more steeply and the big brake time
$t_B$ tends to zero (when $T_*$ gets larger than $T_4$ the
trajectories switch from bundle $f^{III}$ to bundle $f^V$).

Like those of bundles  $f^{III}$   and  $f^V$, the evolutions of
bundle $f^{IV}$ give likewise rise to a big brake and the behavior
of $h(t)$ is qualitatively similar in the two cases. However,
contrary to what happens for bundles    $f^{III}$   and $f^V$, for
each trajectory of bundle   $f^{IV}$    there is a time $t = t_0$
(which depends on the particular trajectory and spans the whole
open half line) for which $s(t_0)= 0$ and hence $\dot{h}(t_0)= 0$.
Therefore, even though the big brake time can get as close to zero
as one likes for the curves of type $IV$ (for such curves, $t_B$
is a decreasing function of the parameter $D$ in Eq. (\ref{3.2}),
with $\lim_{D\to D_c}t_B(D) = \infty$ and $\lim_{D\to
\infty}t_B(D) = 0$) and hence $h(t)$ may decrease to zero very
steeply, there is always some intermediate time at which $h(t)$ is
momentarily stationary. Remark that, as one can see from Fig. 5,
there is nothing peculiar in the behavior of $h(t)$ at the times
when vertices of the rectangle are crossed.

The possibility of cosmological singularities characterized by the
divergence of the second time derivative of the cosmological scale
factor  has also been considered in \cite{Sahni-brake} in the
brane cosmology context.

We conclude this section by another simple example of cosmology
sharing this property. Let us consider the flat Friedmann universe
filled with a perfect fluid with a state equation similar to
(\ref{model4}):
\begin{equation}
p = \frac A\varepsilon
\end{equation}
where  $A$ is positive. We can call this fluid the "anti-Chaplygin
gas". This equation of state arises in the study of the so-called
wiggly strings \cite{Carter,Vilenkin}. The dependence of the
energy density on the cosmological radius is given by
\begin{equation}
\varepsilon = \sqrt{\frac{B}{a^6} -A}, \label{anti-Chap}
\end{equation}
where $B$ is a positive constant. At the beginning of the
cosmological evolution $\varepsilon \sim \sqrt{B}/a^3$, like in
the dust-dominated case. Now there is a maximal value possible for
the cosmological scale
\begin{equation}
a_F = \left(\frac{B}{A}\right)^{1/6}. \label{fin-rad}
\end{equation}
that is attained in a finite cosmic time $t_F$. The behavior of
$a(t)$ in the vicinity of the maximum is the following
\begin{equation}
a(t) \thickapprox a_F - C(t_F - t)^{4/3},\;\;\;\; \ C =
2^{-7/3}3^{5/3}(A B)^{1/6}. \label{anti-Chap1}
\end{equation}
Since $\dot{a}(t_F) = 0$ while $\ddot{a}(t_F) = -\infty$ we are
back into a big brake cosmological singularity.

Thus, a big brake singularity can be found in an elementary
cosmological model (though based on an exotic fluid). The
difference with the tachyonic model is that in the latter there
are evolutions culminating in a big brake which co-exist with
other evolutions giving rise to an  infinite accelerated
expansion. These two types of evolutions correspond to different
classes of initial conditions.

It maybe worth mentioning that also for $k > 0$ one can construct
a scalar field model; indeed the potential displayed in Eq.
(\ref{U2}) is not restricted to $k<0$. But this model has a much
poorer dynamics: all the trajectories tend to the de Sitter
attractive node point.

\section{Final considerations: fate of the universe}
The study of the fate or the future of the universe is  rather
popular nowadays \cite{Dyson}--\cite{Sahni-bounce}, and, as was
predicted more than 20 years ago by F.J. Dyson \cite{Dyson},
``eschatology'' has now become part of the cosmological studies.
This sort of studies include of course also biological aspects of
and the fate of consciousness in the different cosmological
scenarios (see, for example, \cite{Dyson, Freese, Bar-Tip}) but
our goal is much more modest and we shall concentrate on the
purely geometrical facet of the topic.

There are three mainly studied possible scenarios for the future
of the universe, intensively discussed in literature: an infinite
expansion; an expansion followed by a contraction ending in a
``big crunch''; an infinitely bouncing or recycling universe.

The present set of observational data seems to favor the first
scenario: the data  are quite compatible with a flat universe with
a positive cosmological constant.

On the other hand a negative value of the cosmological constant
fits better with string theory (see, e.g.
\cite{Gibbons,Kal-Lin,Kal-Lin1}). The presence of a small negative
cosmological constant could be made compatible with the present
day cosmic acceleration, provided there are some fields or types
of matter responsible for this acceleration. However, in this
context the cosmological radius sooner or later will start
decreasing and the expansion will be followed by a contraction
which would normally lead the universe to a cosmological
singularity of the big crunch type.

The third scenario of an infinitely bouncing or recycling universe
appears to many to be very attractive because it opens an
opportunity to escape the ``frying''  in the big crunch and the
``freezing'' in the infinite expansion case. In this scenario,
during the process of contraction there will be two opportunities:
collapse or bounce with subsequent expansion. The choice of one of
these opportunities depends on the initial conditions and this
dependence has usually a chaotic character
\cite{Page,KKT,KKT1,Corn,KKST}.

Furthermore, one can show \cite{Kal-Lin1} that for every model of
dark energy describing an eternally expanding universe one can
construct many closely related models which describe the present
stage of acceleration of the universe followed by its global
collapse. However, these models corresponding to eternally
expanding and collapsing universes are different and have
different values of their basic parameters.

One interesting feature of our toy tachyon model is that the first
and the second scenario coexist in its context. Depending on
initial conditions, some correspond to eternal expansions of the
universe which approach asymptotically a pure de Sitter regime,
while others end their evolution at the cosmological singularity.

The second distinguishing feature of this model is that the
singularity at which the universe ends its evolution is not a
standard big crunch singularity. Instead, it corresponds to a
finite non zero cosmological radius at which the Hubble parameter
is finite and the deceleration parameter is infinite and has
positive sign. We have called this fate "big brake". The prospect
of hitting the cosmological singularity during expansion has been
also discussed in Ref. \cite{Star-fut}, where the singularity
corresponds to an infinite value of the Hubble variable and the
cosmological radius. This scenario is known as "big rip" or
"phantom cosmology" (see, e.g.
\cite{doomsday,sami-phan,marek,sami-top}).

The third distinguishing aspect of our model is the fact that the
regions of the phase space corresponding to different types of
trajectories are well separated and the dependence of the
cosmology  on the choice of initial conditions is quite regular
(non chaotic). A remark is in order here: the chaoticity of the
classical dynamics hinders the application of the WKB
approximation and, hence, undermines the basis of the majority of
results of quantum cosmology \cite{Corn}. In our model quantum-
cosmological schemes of the traditional type
\cite{Hartle-Hawk,Hawk,Vil,Vil1} can be attempted. The
corresponding wave function should describe a probability
distribution over different initial conditions for the classical
evolution of the universe. The quantum evolution of the universe
in our toy model might be expressed in the language of the
many-worlds interpretation of quantum mechanics
\cite{Everett,DeWitt}. In this framework one can say that the wave
function of the universe describes the quantum birth of the
universe and subsequently the process of the so called
``classicalization'' (see e.g. Ref. \cite{Bar-Kam}). During this
process the wave function splits into different branches or
Everett worlds corresponding to different possible classical
histories of the universe. The peculiarity of our toy model
consists in the fact that some of these branches describe
eternally expanding universes while other branches correspond to
universes which end their evolution hitting the cosmological
singularity.

\section*{Acknowledgements}
A.K. is grateful to CARIPLO Science Foundation and to University
of Insubria for financial support. His work was also partially
supported by the Russian Foundation for Basic Research under the
grant No 02-02-16817 and by the scientific school grant No.
2338.2003.2 of the Russian Ministry of Science and Technology.

\end{document}